\documentclass[11pt]{article}
\pdfoutput=1
\usepackage[totalwidth=480pt, totalheight=680pt]{geometry}

\usepackage[utf8]{inputenc}

\usepackage{hyperref}

\usepackage{amsmath}
\usepackage{amssymb}
\usepackage{latexsym}
\usepackage{bbm}

\usepackage[above,below]{placeins}
\usepackage[title,titletoc]{appendix}

\usepackage{ifpdf}

\ifpdf
\usepackage[pdftex]{graphicx}
\else
\usepackage{graphicx}
\fi

\graphicspath{{curves/}}

\newcommand{\ud}{\mathrm{d}}
\newcommand{\RR}{\mathbb{R}}
\newcommand{\EE}{\mathbb{E}}
\newcommand{\cA}{\mathcal{M}}
\newcommand{\cP}{\mathcal{P}}
\newcommand{\cC}{\mathcal{C}}
\newcommand{\wC}{\widetilde{C}}
\newcommand{\wD}{\widetilde{D}}
\newcommand{\rmin}{\mathrm{min}}
\newcommand{\acosh}{\cosh^{-1}}

\newcommand{\atanh}{\tanh^{-1}}
\newcommand{\atan}{\tan^{-1}}

\title{Asymptotic Implied Volatility at the Second Order with Application to the SABR Model}
\author{Louis Paulot}

\begin{document}

\ifpdf
\DeclareGraphicsExtensions{.pdf, .jpg, .tif}
\else
\DeclareGraphicsExtensions{.eps, .jpg}
\fi

\thispagestyle{empty}


\vspace*{3cm}

\begin{center}
\begin{Large}
\textbf{Asymptotic Implied Volatility at the Second Order
\\
\vspace{3mm}
With Application to the SABR Model}
\end{Large}

\vspace{20mm} {\bf Louis Paulot}

\vspace{10mm}
\emph{Misys
\\
42 rue Washington, 75008 Paris, France}

\vspace{5mm}
{\ttfamily louis.paulot@misys.com}

\vspace{15mm}
June 2009\\
\emph{Revised: August 2014}

\end{center}

\vspace{25mm}
\hrule
\begin{abstract}
We provide a general method to compute a Taylor expansion in time of implied volatility for stochastic volatility models, using a heat kernel expansion. Beyond the order 0 implied volatility which is already known, we compute the first order correction exactly at all strikes from the scalar coefficient of the heat kernel expansion. Furthermore, the first correction in the heat kernel expansion gives the second order correction for implied volatility, which we also give exactly at all strikes. As an application, we compute this asymptotic expansion at order 2 for the SABR model and compare it to the original formula.
\end{abstract}
\hrule

\vspace{\stretch{1}}

\pagebreak

\setcounter{tocdepth}{2}
\tableofcontents

\parskip=4pt


\section{Introduction}

The most known model for pricing derivatives is the Black-Scholes-Merton model, where the underlying is supposed to follow a geometric Brownian motion. Popular extensions include local volatility models and stochastic volatility models. As an example the SABR model \cite{hagan2002msr} combines the local volatility of the CEV model \cite{cox:cev} and a lognormal volatility process. Closed formulas for European options can be obtained for a few models; it is the case of the CEV model or for a stochastic volatility example the Heston model \cite{heston1993cfs}. These are however special cases and there are generally no closed form formulas. Finite difference methods or Monte-Carlo simulations can be used to price derivatives. Approximations have also been computed to achieve faster pricing, especially for calibration processes.

For short maturities, Hagan, Kumar, Lesniewski and Woodward provide an approximation for the implied volatility of the SABR model they introduce \cite{hagan2002msr}. Berestycki, Busca and Florent \cite{berestycki2002aac,berestycki2004civ} and Henry-Labordère \cite{henrylabordere:gai} give general methods to compute short maturity asymptotics of stochastic volatility models. These expansions give the implied volatility at first order in maturity. In addition some quantities are approximated by their value at the money, which can produce errors in the wings of the distributions.

In this paper, we leverage on the heat kernel methods introduced for the study of stochastic processes by Varadhan \cite{varadhan1967dps,varadhan1967bfs} and used for the SABR model in \cite{hagan2001pds} and \cite{henrylabordere:gai}. Using the heat kernel expansion of DeWitt \cite{dewitt1965dtg}, we provide a method to compute exactly a Taylor expansion of the implied volatility at all strikes. The stochastic volatility diffusion is formulated as a diffusion on a Riemannian manifold. The geodesic distance gives the implied volatility at null maturity. The multiplicative factor of the heat kernel expansion provides the first order (in time) correction to implied volatility. The first corrective term of the heat kernel is translated into the second order correction to implied volatility and similarly for higher order corrections. We perform a detailed computation up to order 2 of the Taylor expansion in time of implied volatility, without other approximations.

More generally, our method can be used to approximate a stochastic volatility model by an other model for which a closed form solution exists, with an implied parameter computed as a Taylor expansion.

As an application, we compute the asymptotic SABR volatility at order 2 and compare it to finite difference method results and to the original SABR expansion.

Our results can be useful for pricing short maturities options or even long maturities options with low volatility of volatility. When the approximation is not valid, a numerical method such as a finite difference method (FDM) has to be used. When our approximation is valid, it gives much faster results. At very short maturities, the prices are even more precise. Calibration at short maturities appears to be more stable using this approximation.

In section~\ref{sec:diffusion} we recast the financial model in physical and geometric terms and fix our conventions. In section~\ref{sec:asymptotic} we use a heat kernel expansion to compute a short maturity expansion of Black or more generally CEV implied volatility. Finally in section~\ref{sec:SABR} we apply the method to the SABR model and compare the results to FDM and to the original formula.

\section{Diffusion equation in covariant form}
\label{sec:diffusion}

A stochastic volatility model for some asset with pure diffusion (no jumps) is described by two risk-neutral processes: the asset price $S$ and a variable $V$ which describes the stochastic part of volatility. In the Heston model $V$ would be the variance whereas in the SABR model it is a factor of volatility. The diffusion is given by the stochastic differential equations
\begin{eqnarray}
\ud S &=& \mu_S(S) \ud t + \sigma_S(S,V) \ud W_1
\nonumber \\
\ud V &=& \mu_V(V) \ud t + \sigma_V(V) \ud W_2
\label{eq-stoch}
\end{eqnarray}
where $\ud W_1$ and $\ud W_2$ are two standard Brownian processes with correlation $\rho$. The dependence of parameters in variables $S$ and $V$ we have written is the more common, it may be more general with all parameters depending on both variables.

Stochastic volatility models can be seen as diffusions on a Riemann surface. More precisely, prices of securities are sections of a line bundle over this Riemann surface which are solutions of a diffusion (or heat) equation.

A introduction to this subject and its applications to finance can be found in~\cite{henrylabordere2008aga}. We present here the formalism and define all quantities we use in order to set our conventions.

\subsection{Diffusion equation}

Let us consider a general model with $n$ state variables $X^i(t)$ (which will be the spot and the volatility) which follow a pure diffusion process, without jumps. For simplicity we consider a European payoff of some maturity $T$. The price $P(X(t),t)$ of such a payoff is the solution of a diffusion equation
\begin{equation}
- \partial_t P = \mu^i \partial_i P + \frac{1}{2} \Sigma^{ij} \partial_i \partial_j P - r P
\label{eq-diff}
\end{equation}
where $\Sigma^{ij}$ is the covariance matrix, $\mu_i$ the drifts and $r$ the numéraire rate. All coefficients can depend on state variables $X^i(t)$ and time $t$. Unless explicitly staten, we adopt Einstein sum convention: repeated indices are summed. The price of the European option is given by the solution of this equation with terminal boundary condition at maturity $T$ given by the payoff.

The covariance matrix $\Sigma^{ij}$ can be seen geometrically as the inverse $g^{ij} = \Sigma^{ij}$ of a metric $g_{ij}$ on the space of variables. The diffusion equation describes the diffusion over a Riemannian manifold: the state of variables endowed with the metric $g_{ij} = \left(\Sigma^{-1}\right)_{ij}$.

\

\noindent{\bf Examples}
\begin{enumerate}
\item
The Black-Scholes equation in the monetary account numéraire with volatility $\sigma$, risk free rate $r$ and dividend yield $q$ reads
\begin{equation*}
- \partial_t P = (r-q)S \partial_S P + \frac{1}{2} \sigma^2 S^2 \partial_S \partial_S P - r P
\end{equation*}
This is equation \eqref{eq-diff} with $\mu^S = (r-q)S$, $\Sigma^{SS} = \sigma^2 S^2$ and $r=r$.
\item
For the stochastic volatility model described by equations \eqref{eq-stoch}, $\mu^i$ is a two-dimensional vector
\begin{equation*}
\mu = \left(
\begin{array}{c}
\mu_S
\\
\mu_V
\end{array}
\right)
\rlap{\ .}
\end{equation*}
The covariance matrix $\Sigma^{ij}$ is
\begin{equation*}
\Sigma = \left(
\begin{array}{cc}
{\sigma_S}^2 & \rho \sigma_S \sigma_V
\\
\rho \sigma_S \sigma_V & {\sigma_V}^2
\end{array}
\right)
\rlap{\ .}
\end{equation*}
\end{enumerate}

In what follows we restrict ourselves to the case of time-homogeneous models: there is no explicit time-dependence in parameters. The generalization to time-dependent cases is not difficult.

\subsection{Gauge structure}
\label{subsec-gauge}

There are several \emph{gauge} transformations which are natural for such systems:
\begin{enumerate}
\item Change of numéraire:
\begin{equation*}
P(X,t) \ \longrightarrow \ \frac{P(X,t)}{\Phi(X,t)}
\end{equation*}
where $\Phi(X,t)$ is the price of a security which is always nonzero. Mathematically it is a real function which is positive everywhere, that we denote thus by $\Phi(X,t) = e^{-\phi(X,t)}$.
\item Change of variables
\begin{equation*}
X \ \longrightarrow \ X'(X) \rlap{\ .}
\end{equation*}
\end{enumerate}

The natural way to handle a system with gauge freedom is to introduce covariant derivatives. The coordinate freedom is handled through the \emph{Levi-Civita} connection which acts respectively on scalars, vectors and 1-forms as
\begin{eqnarray*}
D _i f &=& \partial_i f
\\
D _i f^j &=& \partial_i f^j + \Gamma^j_{ik} f^k
\\
D _i f_j &=& \partial_i f_j - \Gamma^k_{ij} f_k
\end{eqnarray*}
where $\Gamma^j_{ik}$ are the \emph{Christoffel symbols}. The action on tensors with more indices is obtained by acting on all indices with the Christoffel symbols. Christoffel symbols can be computed from the metric as
\begin{equation*}
\Gamma^k_{ij} = \frac{1}{2} g^{kl} \left( \partial_i g_{lj} + \partial_j g_{il} - \partial_l g_{ij} \right) \rlap{\ .}
\end{equation*}
A fundamental property of the Levi-Civita connection is the covariance of the metric:
\begin{equation*}
D_i g_{jk} = 0 \rlap{\ .}
\end{equation*}
The metric is used to transforms vectors into 1-forms and conversely, \emph{i.e.} lowering or raising indices:
\begin{eqnarray*}
A^i &=& g^{ij} A_j
\\
A_i &=& g_{ij} A^j \rlap{\ .}
\end{eqnarray*}

The numéraire gauge freedom is handled through a line bundle $\mathcal{L}$ (\emph{i.e.} with sections in $\mathbb{R}$).  Geometrically, $P$ is a section of $\mathcal{L}$. A $\mathbb{R}$-valued connection\footnote{
This connection is similar to the connection which described the electromagnetic potential, except that the fibre of the gauge bundle is $\mathbb{R}$ instead of $U(1)$. This causes a difference of a factor $i$ in equations.} is defined with spatial and time components given by a 1-form $A_i$ and a scalar\footnote{
There is a breaking of symmetry between time and spatial directions. The diffusion equation can be seen as a non-relativistic limit of a pure wave equation in imaginary time.}
$Q$:
\begin{eqnarray*}
\nabla_i P &=& (D_i - A_i) P
\\
\nabla_t P &=& (\partial_t - Q) P \rlap{\ .}
\end{eqnarray*}
Under the change of numéraire
\begin{eqnarray*}
P  & \longrightarrow & e^{\phi(X,t)} P
\end{eqnarray*}
these operators are covariant,
\begin{eqnarray*}
\nabla_i P & \longrightarrow & e^{\phi(X,t)} \nabla_i P
\\
\nabla_t P & \longrightarrow & e^{\phi(X,t)} \nabla_t P \rlap{\ ,}
\end{eqnarray*}
provided that $A_i$ and $Q$ are shifted as
\begin{eqnarray*}
A_i & \longrightarrow & A_i - \partial_i \phi
\\
Q & \longrightarrow & Q - \partial_t \phi \rlap{\ .}
\end{eqnarray*}

Using these connections, the diffusion equation \eqref{eq-diff} can be rewritten as
\begin{equation}
-\nabla_t P = \frac{1}{2} \nabla^i \nabla_i P \rlap{\ .}
\label{eq-diff-cov}
\end{equation}

Identifying terms between equations \eqref{eq-diff} and \eqref{eq-diff-cov}, the $\mathbb{R}$ connection must be
\begin{eqnarray}
A_i &=& g_{ij} \left( - \frac{1}{2} \Gamma^j_{kl} g^{kl} - \mu^j \right)
\label{eq-A}
\\
Q &=& -\frac{1}{2} g^{ij} \left( \partial_i A_j - A_i A_j - \Gamma^k_{ij} A_k \right)  + r \rlap{\ .}
\label{eq-Q}
\end{eqnarray}
In addition with
\begin{equation*}
g^{ij} = \Sigma^{ij}
\end{equation*}
this translates the set of financial parameters into geometrical quantities.

\subsection{Kolmogorov forward equation}

The Kolmogorov backward equation \eqref{eq-diff-cov} leads to a dual Kolmogorov forward equation.

We suppose that all prices are expressed with respect to a numéraire which is a traded asset that does not pay any coupon or dividend. The price of the numéraire security itself is identically 1; this reads mathematically
\begin{equation*}
-\nabla_t 1 = \frac{1}{2} \nabla^i \nabla_i 1
\end{equation*}

If $p(X,t)$ is a risk-neutral probability density to get in state $X$ at time $t$ starting from state $X_0$ at time 0, then the price of a European payoff of maturity $T \geq t $ can be written as
\begin{equation*}
P(X_0,0) = \int \ud X p(X,t) P(X,t)
\end{equation*}
As $t$ does not appear on the left-hand side, the derivative of the integral with respect to $t$ must vanish.
\begin{equation*}
\int \ud X \partial_t (p(X,t) P(X,t)) = 0 \rlap{\ .}
\end{equation*}
We define an action of the gauge group on $p$ with a plus sign instead of a minus sign when acting on $P$:
\begin{eqnarray*}
\nabla_i p &=& (D_i + A_i) p
\\
\nabla_t p &=& (\partial_t + Q) p \rlap{\ .}
\end{eqnarray*}
This means that they $p$ and $P$ have opposite charges under the numéraire $\RR$ gauge group, such that $pP$ is neutral and $\nabla_t(pP) = \partial_t(pP)$. We have thus
\begin{equation*}
\int \ud X \left( \nabla_t p(X,t) P(X,t) + p(X,t) \nabla_t P(X,t) \right) = 0
\end{equation*}
Using equation \eqref{eq-diff-cov} for $\nabla_t P(X,t)$ and integrating by part on the spatial directions, this equation becomes
\begin{equation}
\int \ud X \left( \nabla_t p(X,t) - \frac{1}{2} \nabla^i \nabla_i p(X,t) \right) P(X,t) = 0
\label{eq-kolmo-interm}
\end{equation}
This equation will be automatically satisfied if
\begin{equation}
\nabla_t p = \frac{1}{2} \nabla^i \nabla_i p \rlap{\ .}
\label{eq-kolmo}
\end{equation}
Moreover, if the market is complete equation \eqref{eq-kolmo-interm} must be true for all functions $P(\cdot,t)$ which imposes equation \eqref{eq-kolmo}.
This is the Kolmogorov forward equation, written in a covariant way.

It should be noted that $p(X,t)$ is a density, which means that the Levi-Civita connection does not reduce to a partial derivative as would be the case for a scalar. More precisely, the transition probability $p(X_0,0;X,t)$ has value in $\mathcal{L} \boxtimes \mathcal{L}^* \otimes \wedge^d (T^*M) $. Numéraire gauge tranformations associated with the line bundle $\mathcal{L}$ acting on $p$ gives the well-known change of measure which are usually obtained from the Girsanov formula.

\section{Asymptotic implied volatility}
\label{sec:asymptotic}

We consider a stochastic volatility model where the variable is a forward price or rate $F$ with a volatility variable $V$:
\begin{eqnarray*}
\ud F &=& \sigma_F(F,V) \ud W_1
\nonumber \\
\ud V &=& \mu_V(V) \ud t + \sigma_V(V) \ud W_2
\end{eqnarray*}
with $ \langle \ud W_1 \ud W_2\rangle = \rho \ud t$.

\

Our computation of an asymptotic expansion at short time of implied volatility at strike $K$ involves four steps:
\begin{enumerate}
\item Compute an asymptotic value of the transition probability from initial state $F_0$, $V_0$ at time 0 to $K$, $V$ at time $t$ using a heat kernel expansion;
\item Compute $\EE\!\left[\sigma_F^2(K,V) \delta(F-K)\right]$ using a saddle point method;
\item Integrate over time to compute the time value;
\item Compare to the same formula for the Black-Scholes model to extract the implied volatility.
\end{enumerate}

\subsection{Heat kernel expansion}

In order to keep exposition as simpler and clear as possible, we will skip here technical details and refer the reader to \cite{dewitt1965dtg} or \cite{vassilevich2003hep} for mathematically precise statements\footnote{In finance we will usually consider noncompact manifolds, possibly with boundaries as in the SABR model for $0<\beta<1$.}.

At short time the solution of equation \eqref{eq-kolmo} with initial condition $p(X,0) = \delta(X-X_0)$ is asymptotically given by a heat kernel expansion\footnote{
Using Feynman path integral, the solution to equation \eqref{eq-kolmo} can be written up to some normalization factor as
\begin{equation*}
p(X,t) \propto \int [DX]  e^{\displaystyle - \int_0^t \ud t \left( \frac{1}{2} g_{ij} \dot X^i \dot X^j + A_i \dot X^i + Q \right)}
\end{equation*}
where $[DX]$ means integrating over all path $X(s)$ going from $X(0) = X_0$ to $X(t) = X$. The normalization factor is the inverse of the same quantity with the integral computed over all paths with starting point $X_0$, so that the total probability is 1.
It is generally not possible to compute this integral exactly. However it gives some hints on the asymptotic solution at short time: the solution will be dominated by the path corresponding to the minimal value of the integrand inside the exponential, which will be close to the geodesic path.
}
\begin{equation}
p(X,t) = \frac{\sqrt{g(X)}}{(2 \pi t)^{n/2}} \sqrt{\Delta(X_0,X)} \cP(X_0,X) e^{\displaystyle -\frac{d^2(X_0,X)}{2t}} \sum_{k \geq 0} a_k(X_0,X) t^k \rlap{\ .}
\label{eq-heat-expansion}
\end{equation}
$g(X)$ is the determinant of the metric at point $X$:
\begin{equation*}
g = \mathrm{Det}(g_{ij}) \rlap{\ .}
\end{equation*}
$d(X_0,X)$ is the geodesic distance between the starting point $X_0$ and the end point $X$, this is the minimal distance between $X_0$ and $X$. It can also be written as
\begin{equation*}
\frac{d^2(X_0,X)}{t} = \min_{X(s)} \int_0^t \ud t \, g_{ij} \dot X^i \dot X^j
\end{equation*}
where the minimum is taken on all paths going from $X(0) = X_0$ to $X(t) = X$. (This is independent of $t$.) We denote by $\cC$ this geodesic path.
$\Delta(X_0,X)$ is the Van Vleck--Morette determinant
\begin{equation*}
\Delta(X_0,X) = \frac{\displaystyle \mathrm{Det}\!\left( - \frac{1}{2} \frac{\partial^2 d^2(X_0,X)}{\partial X^i \partial X_0^j} \right)}{\sqrt{g(X_0) g(X)}}
\end{equation*}
$\cP(X_0,X)$ is the parallel transport along the geodesic with respect to the $\RR$ connection. It is such that its covariant derivative along the geodesic path is null:
\begin{equation*}
\cP(X_0,X) = e^{\displaystyle - \int_\cC A_i \dot X^i \ud t} = e^{\displaystyle - \int_\cC A_i \ud X^i}
\end{equation*}
where the integral is computed on the geodesic path $\cC$.
Finally, $a_i(X_0,X)$ are functions which are defined recursively with
\begin{equation*}
a_0 = 1
\end{equation*}
and $a_i$'s satisfy the differential equations
\begin{equation*}
(k + (\nabla^i d) d \nabla_i)a_k = \cP^{-1} \Delta^{-1/2} \left(\frac{1}{2}\nabla^i\nabla_i - Q \right) \cP \Delta^{1/2} a_{k-1} \rlap{\ .}
\end{equation*}
Along a given geodesic curve parameterized by its geodesic distance from $X_0$, this equation reads
\begin{equation*}
(k + d \partial_d )a_k = \cP^{-1} \Delta^{-1/2} \left(\frac{1}{2}\nabla^i\nabla_i - Q \right) \cP \Delta^{1/2} a_{k-1}
\end{equation*}
which can be integrated as
\begin{equation*}
a_k = \frac{1}{d^k} \int_0^d ds \, s^{k-1} \cP^{-1} \Delta^{-1/2} \left(\frac{1}{2}\nabla^i\nabla_i - Q \right) \cP \Delta^{1/2} a_{k-1} \rlap{\ .}
\end{equation*}

Functions $a_k$ are sections of a $\mathcal{L} \boxtimes \mathcal{L}^*$ bundle. The parallel transport $\cP$ and the connexion with respect to the numéraire gauge group act on the second factor of this external product. (The first factor is related to the numéraire at $t=0$.) Also note that $\frac{p(X,t)}{\sqrt{g(X)}}$ is a scalar with respect to the Levi-Civita connection. 

In order to produce a first order expansion of the implied volatility, only the common multiplicative factor of expansion~\eqref{eq-heat-expansion} is needed. In order to compute a second order term for the implied volatility, we will also make use of the first corrective term $a_1 t$ with
\begin{equation}
a_1 = \frac{1}{d} \int_0^d ds \, \cP^{-1} \Delta^{-1/2} \left(\frac{1}{2}\nabla^i\nabla_i - Q \right) \cP \Delta^{1/2} \rlap{\ .}
\label{eq-a1}
\end{equation}

\subsection{Expected variance}

We now compute $\EE\!\left[\sigma_F^2(F_t,V_t) \delta(F_t-K)\right]$. This quantity can be written as an integral over the terminal volatility variable $V$:
\begin{equation*}
\EE\!\left[\sigma_F^2(F_t,V_t) \delta(F_t-K)\right] = \int \ud V\, \sigma_F^2(K,V) p(K,V;t)
\end{equation*}
where $p(F,V;t)$ is given by the heat kernel expansion \eqref{eq-heat-expansion} with $X = \left( \begin{array}{c} F \\ V \end{array} \right)$ and $n=2$.
The integrand can be written as
\begin{equation}
\sigma_F^2(K,V) p(K,V;t) = \frac{1}{2\pi t} e^{\displaystyle -\frac{B}{t}-C-Dt + o(t)}
\label{eq-saddle-integrand}
\end{equation}
with
\begin{eqnarray}
B &=& \frac{1}{2} d(F_0,V_0;K,V)^2
\label{eq-saddle-integrand-B}
\\
C &=& -2\ln(\sigma_F(K,V)) - \frac{1}{2} \left[ \ln( g(K,V)) + \ln(\Delta(F_0,V_0;K,V)) \right] + \cA(K,V)
\label{eq-saddle-integrand-C}
\\
D &=& - a_1(K,V)
\label{eq-saddle-integrand-D}
\end{eqnarray}
where $\cA$ is the integral of the $\RR$ connection
\begin{equation*}
\cA(K,V) = -\ln(\cP(K,V)) = \int_\cC A_i \ud X^i
\end{equation*}
on $\cC$, the geodesic curve joining $(F_0,V_0)$ to $(K,V)$,
and $a_1$ is given in equation \eqref{eq-a1} as an integral over the geodesic path .

The integral over \eqref{eq-saddle-integrand} will be dominated at short time by the $B$ term. More precisely, it will be dominated by the volatility $V_\rmin$ which minimizes $B(K,V) = \frac{1}{2}d(F_0,V_0;K,V)^2$. This is the final volatility which minimizes the distance between the initial conditions and the strike $K$. Expanding all functions in the neighborhood of $V_\rmin$, where $B'(V_\rmin) = 0$, the integrand is
\begin{multline*}
\sigma_F^2(K,V) p(K,V;t) = \frac{1}{2\pi t} e^{\displaystyle -\frac{B}{t}-C-Dt} e^{\displaystyle -\frac{B''}{2 t}  \delta V^2 }
\\
\left[ 1 - \frac{1}{2} \left( C'' - {C'}^2 \right) \delta V^2 - \left( \frac{1}{24}B^{(4)} - \frac{1}{6} B^{(3)} C' \right) \frac{\delta V^4}{t} + \frac{1}{72} {B^{(3)}}^2 \, \frac{\delta V^6}{t^2} + o(t) + \mathrm{odd\ terms} \right]
\end{multline*}
where derivatives are with respect to $V$, all functions $B$, $C$, $D$ and their derivatives are taken at $(K,V_\rmin)$ and $\delta V = V-V_\rmin$. When writing $o(t)$, we have anticipated that after integration $\delta V^2 \sim \frac{1}{t}$. We have also anticipated that odd terms in $\delta V$ will not give contributions to the integral.

Integrating over $\delta V$, and using that the first even moments of the standard normal distribution are $M_2 = 1$, $M_4 = 3$ and $M_6 = 15$, we get for the integral
\begin{multline*}
\EE\!\left[\sigma_F^2(F_t,V_t) \delta(F_t-K)\right] = \frac{1}{\sqrt{2\pi t B''}} e^{\displaystyle -\frac{B}{t}-C-Dt}
\\
\left[ 1 - \frac{1}{2} \left( C'' - {C'}^2 \right) \frac{t}{B''} - \left( \frac{1}{24}B^{(4)} - \frac{1}{6} B^{(3)} C' \right) \frac{3t}{{B''}^2} + \frac{1}{72} {B^{(3)}}^2 \, \frac{15 t}{{B''}^3} + o(t) \right]
\end{multline*}
This can be rewritten as
\begin{equation}
\EE\!\left[\sigma_F^2(F_t,V_t) \delta(F_t-K)\right] = \frac{1}{\sqrt{2\pi t}} e^{\displaystyle -\frac{B}{t}-\wC-\wD t +o(t) }
\label{eq-expected}
\end{equation}
with
\begin{eqnarray}
\wC &=& C + \frac{1}{2} \ln(B'')
\label{eq-expected-C}
\\
\wD &=& D + \frac{1}{2 B''}\left[ C'' - {C'}^2 + \frac{1}{4} \frac{B^{(4)}}{B''} - \frac{B^{(3)}}{B''} C' - \frac{5}{12} \left( \frac{B^{(3)}}{B''}\right)^2 \right]
\nonumber
\\
&=&  D + \frac{1}{2 B''}\left[ \wC'' - \wC'^2 - \frac{1}{4} \frac{B^{(4)}}{B''} + \frac{1}{3} \left( \frac{B^{(3)}}{B''}\right)^2 \right]
\label{eq-expected-D}
\end{eqnarray}
where all derivatives are with respect to $V$ and all functions and their derivatives are taken at $(K,V_\rmin)$.

\subsection{Time value}

The price of a Call of maturity $T$ and strike $K$ can be written as the payoff integrated against the risk-neutral distribution:
\begin{equation*}
\mathrm{Call}(K,T) = e^{-rT} \int \ud F \, (F-K)^+ p(F,T)
\end{equation*}
where $p(F,t)$ is the marginal probability density of $F_t$.
This can be written also as a double integral over forward and time as
\begin{equation}
\mathrm{Call}(K,T) = e^{-rT} \left[ (F_0-K)^+ + \int \ud F \int_0^T \ud t \, (F-K)^+ \partial_t p(F,t) \right] \rlap{\ .}
\label{eq-call-interm}
\end{equation}
As $F_t$ is a forward, it is a driftless process and the Kolmogorov forward equation reduces to
\begin{equation}
\partial_t p(F,t) = \frac{1}{2} \partial_F^2 \left( \sigma_{loc}^2(F,t) p(F,t)\right)
\label{eq-call-kolmo}
\end{equation}
where $\sigma_{loc}^2(F,t)$ is the local (normal) volatility
\begin{equation*}
\sigma_{loc}^2(F,t) = \EE\!\left[\sigma_F^2(F_t,V_t) \mid F_t=F \right] = \frac{\EE\!\left[\sigma_F^2(F_t,V_t) \delta(F_t-F)\right]}{p(F,t)} \rlap{\ .}
\end{equation*}
Plugging the Kolmogorov equation \eqref{eq-call-kolmo} in equation \eqref{eq-call-interm} and integrating twice by part on the $F$ variable, the Call price is finally obtained as an integral over time at strike $K$:
\begin{equation}
\mathrm{Call}(K,T) = e^{-rT} \left[ (F_0-K)^+ + \frac{1}{2} \int_0^T \ud t \, \EE\!\left[\sigma_F^2(F_t,V_t) \delta(F_t-K)\right] \right] \rlap{\ .}
\label{eq-call}
\end{equation}

Using expression \eqref{eq-expected} for the integrand, the integral over time can be computed:
\begin{multline*}
\frac{1}{2} \int_0^T \ud t \, \EE\!\left[\sigma_F^2(F_t,V_t) \delta(F_t-K)\right] =
\\
\frac{1}{\sqrt{2}} \, e^{\displaystyle-\wC} \left[ \sqrt{\frac{t}{\pi}} e^{\displaystyle -\frac{B}{t}} - \sqrt{B} \, \mathrm{erfc}\!\left( \sqrt{\frac{B}{t}} \right)
- \frac{\wD}{3} \left(\sqrt{\frac{t}{\pi}} (t-2B) e^{\displaystyle -\frac{B}{t}} + 2 B^{3/2} \, \mathrm{erfc}\!\left( \sqrt{\frac{B}{t}} \right)  \right) \right]
\\
+ o\!\left( t^{5/2} \, e^{\displaystyle -\frac{B}{t}}  \right)
\end{multline*}
where $\mathrm{erfc}$ is the complementary error function, equal to the cumulative of the standard normal distribution up to $\sqrt{2}$ factors:
\begin{equation*}
\mathrm{erfc}(x) = \frac{2}{\sqrt{\pi}}\int_x^{+\infty}\ud y \, e^{-y^2} = 2\,  \mathcal{N}\!\left(-\sqrt{2} x \right) \rlap{\ .}
\end{equation*}
The asymptotic expansion of this function at $+\infty$
\begin{equation*}
\mathrm{erfc}(x) = \frac{\displaystyle e^{-x^2}}{x \sqrt{\pi}} \left( 1 - \frac{1}{2 x^2} + \frac{3}{4 x^4} + o\!\left( \frac{1}{x^4} \right) \right)
\end{equation*}
with $x=\sqrt{\frac{B}{t}}$
gives the asymptotic expansion of the time value:
\begin{equation}
\frac{1}{2} \int_0^T \ud t \, \EE\!\left[\sigma_F^2(F_t,V_t) \delta(F_t-K)\right] =
\frac{T^{3/2}}{2 \sqrt{2\pi}} \, e^{\displaystyle -\frac{B}{T}-\wC-\ln(B)-\wD \, T - \frac{3}{2B} T + o(T) } \rlap{\ .}
\label{eq-time-value}
\end{equation}

\subsection{Implied volatility}
\label{subsec:BS}

The final step consists in computing the same expansion for the Black--Scholes model, which is simpler as there is no stochastic volatility to be integrated.
The metric is given by the inverse of the variance:
\begin{equation*}
g_{FF} = \frac{1}{\sigma^2 F^2} \rlap{\ .}
\end{equation*}
The Christoffel symbol is therefore
\begin{equation*}
\Gamma^F_{FF} = - \frac{1}{F} \rlap{\ .}
\end{equation*}
The $\RR$-connection components are computed using \eqref{eq-A} and \eqref{eq-Q}:
\begin{eqnarray*}
A_F &=& \frac{1}{2F}
\\
Q &=& \frac{\sigma^2}{8} \rlap{\ .}
\end{eqnarray*}
The geodesic distance is
\begin{equation*}
d(F_0,K) = \left\vert \int_{F_0}^K \frac{\ud F}{\sigma F} \right\vert =  \frac{1}{\sigma} \left\vert \ln\frac{K}{F_0} \right\vert \rlap{\ .}
\end{equation*}
The Van Vleck--Morette determinant is simply
\begin{equation*}
\Delta(F_0,K) = 1 \rlap{\ .}
\end{equation*}
The parallel transport is
\begin{equation*}
\cP(F_0,K) = e^{\displaystyle - \int_{F_0}^K \ud F \, A_F } =  \sqrt{\frac{F}{K}} \rlap{\ .}
\end{equation*}
Putting all these elements together, the heat kernel expansion of $p(K,t)$ is according to \eqref{eq-heat-expansion}
\begin{equation*}
p(K,t) = \frac{1}{\sigma K \sqrt{2\pi t}} \sqrt{\frac{F}{K}} e^{\displaystyle - \frac{\ln^2\frac{K}{F}}{2 \sigma^2 t}} \left(1-\frac{\sigma^2}{8} t + o(t) \right)  \rlap{\ .}
\end{equation*}
Multiplying by the local variance, we get
\begin{equation}
\EE\!\left[\sigma_F^2(K) \delta(F-K)\right] = \sigma^2 K^2 p(K,t) = \frac{1}{\sqrt{2\pi t}} e^{\displaystyle -\frac{B_\mathrm{BS}}{t}-\wC_\mathrm{BS}-\wD_\mathrm{BS} t +o(t) }
\label{eq-BS}
\end{equation}
with
\begin{eqnarray*}
B_\mathrm{BS} &=& \frac{1}{2 \sigma^2} \ln^2\frac{K}{F_0}
\\
\wC_\mathrm{BS} &=& -\ln(\sigma) - \frac{1}{2} \ln(K F_0)
\\
\wD_\mathrm{BS} &=& \frac{\sigma^2}{8} \rlap{\ .}
\end{eqnarray*}
(In fact formula \eqref{eq-BS} is exact: there is no $o(t)$ correction and it can be integrated exactly to get the Black--Scholes formula.)

Writing equation \eqref{eq-time-value} for both the stochastic volatility model and the Black-Scholes model, the implied volatility is such that both quantities are equal:
\begin{equation}
\frac{B}{T}+\wC+\ln(B)+\wD T + \frac{3}{2B} T = \frac{B_\mathrm{BS}}{T}+\wC_\mathrm{BS}+\ln(B_\mathrm{BS})+\wD_\mathrm{BS}T + \frac{3}{2B_\mathrm{BS}} T + o(T) \rlap{\ .}
\label{eq-implied-interm}
\end{equation}
Expanding the implied volatility $\sigma$ as a Taylor expansion
\begin{equation*}
\sigma(K,T) = \sigma_0(K) + \sigma_1(K) T + \sigma_2(K) T^2 + o(T^2)
\end{equation*}
and plugging this into equation \eqref{eq-implied-interm} on the Black--Scholes side, we get
\begin{multline*}
\frac{1}{2 \sigma_0^2 T} \ln^2\!\left(\frac{K}{F_0}\right) \left( 1 - 2 \frac{\sigma_1}{\sigma_0} \, T - 2 \frac{\sigma_2}{\sigma_0} \, T^2 + 3 \left(\frac{\sigma_1}{\sigma_0}\right)^2 T^2 \right)
-\ln(\sigma_0) - \frac{\sigma_1}{\sigma_0} \, T - \frac{1}{2} \ln(K F_0)
\\
+\ln\!\left(\frac{1}{2 \sigma_0^2} \ln^2\frac{K}{F_0}\right) - 2 \frac{\sigma_1}{\sigma_0} \, T
+ \frac{\sigma_0^2}{8} \, T
+ \frac{3 \sigma_0^2 }{\ln^2\frac{K}{F_0}} \, T
= \frac{B}{T}+\wC+\ln(B)+\wD \, T + \frac{3}{2B} \, T + o(T) \rlap{\ .}
\end{multline*}
Coefficients must be equal at each order in $T$, which gives our final expansion of the implied volatility.

Power $-1$ gives the order 0 implied volatility
\begin{equation}
\sigma_0 = \frac{\left\vert \ln\frac{K}{F_0} \right\vert}{\sqrt{2 B}} = \frac{\left\vert \ln\frac{K}{F_0} \right\vert}{d(F_0,V_0;K,V_\rmin)}
\label{eq-order0}
\end{equation}
which was already obtained in \cite{berestycki2002aac} and \cite{henrylabordere:gai}.

The first order correction is extracted from the constant term:
\begin{equation}
\frac{\sigma_1}{\sigma_0} = - \frac{\wC + \ln\!\left(\sigma_0 \sqrt{K F_0}\right)}{2 B} \rlap{\ .}
\label{eq-order1}
\end{equation}

Finally the $O(T)$ term gives the second order correction:
\begin{equation}
\frac{\sigma_2}{\sigma_0} = \frac{3}{2} \left(\frac{\sigma_1}{\sigma_0}\right)^2 - \frac{1}{2B} \left( \wD + 3\frac{\sigma_1}{\sigma_0} - \frac{\sigma_0^2}{8}\right) \rlap{\ .}
\label{eq-order2}
\end{equation}

This gives our final result as the implied volatility expansion
\begin{equation}
\sigma = \sigma_0 \left( 1 + \frac{\sigma_1}{\sigma_0} \, T + \frac{\sigma_2}{\sigma_0} \, T^2 + o(T^2) \right) \rlap{\ .}
\end{equation}
We stress that this result is exact in strike: for a given strike, we have computed exactly the three first coefficients of the Taylor expansion. Moreover, contrary to other expansions, the order 1 expansion is extracted from the order 0 expansion of the probability. This technique allows us to extract a second order term for the implied volatility from the order 1 term in the probability expansion. This method can be used to compute the Taylor expansion of implied volatility up to any order, although the computation becomes more complicated and involves integrals of increasing dimension: the $a_k$ coefficient of the heat kernel expansion involves $k+1$ integrals.

\subsection{At the money}

The computation we have performed makes the implicit hypothesis that we are not exactly at the money: $K \neq F_0$. Otherwise, the dominant term in the exponential would vanish and we could not use the asymptotic expansion of the erfc function at infinity. Precisely at the money, we should use instead a Taylor expansion in 0. As the implied volatility surface is smooth, we just take the limit of formulas \eqref{eq-order0}, \eqref{eq-order1} and \eqref{eq-order2} at $K \rightarrow F_0$. If we perform instead the Taylor expansion of the erfc function at 0, we find only the two first orders
\begin{eqnarray*}
\sigma_0(F_0) &=& \frac{e^{-\wC(F_0)}}{F_0}
\\
\frac{\sigma_1}{\sigma_0}(F_0) &=& \frac{1}{3} \left( \frac{\sigma_0^2}{8} - \wD(F_0) \right) \rlap{\ .}
\end{eqnarray*}
Careful Taylor expansions of all quantities at the money can be used to check that this is indeed the limit of equations \eqref{eq-order0} and \eqref{eq-order1}. Moreover, it can be seen that the existence of these limit are conditions for formulas \eqref{eq-order1} and \eqref{eq-order2} to be convergent, as $B$ goes to 0 at the money (at order 2 in the geodesic distance, which means that the numerators must in fact vanish at order 2).

\subsection{CEV volatility}
\label{subsec:cev-theo}

Instead of Black volatility, the asymptotic expansion can be computed for other local volatility models. Without stochastic volatility, the SABR model reduces to the CEV model. The local volatility part of the model is thus taken into account exactly without introducing approximation besides the stochastic corrections. In view of our application to the SABR model, we will compute here a CEV implied volatility. There are closed formulas for this model, involving Bessel functions. This implied volatility can therefore be used in the CEV pricing formula in order to get the price of the option.

For a CEV model with parameter $\beta_0$ and volatility factor $\sigma$, such that
\begin{equation*}
\ud F = \sigma F^{\beta_0} \ud W \rlap{\ ,}
\end{equation*}
the function $B$, $\wC$ and $\wD$ are
\begin{eqnarray*}
B_0 &=& \frac{1}{2 \sigma^2} \ln^2(q_0)
\\
\wC_0 &=& -\ln(\sigma) - \frac{1}{2} \beta_0 \ln(K F_0)
\\
\wD_0 &=& \frac{\beta_0 (2-\beta_0) \sigma^2}{8 K^{1-\beta_0} F_0^{1-\beta_0}}
\end{eqnarray*}
with
\begin{equation*}
q_0 = \left\{ \begin{array}{ll}
\displaystyle \frac{K^{1-\beta_0}-F_0^{1-\beta_0}}{1-\beta_0} & \beta_0 <1
\\
\displaystyle \ln\!\left(\frac{K}{F_0}\right) & \beta_0 = 1 \rlap{\ .}
\end{array} \right.
\end{equation*}

Formulas \eqref{eq-order0}, \eqref{eq-order1} and \eqref{eq-order2} are modified as follows.
\begin{equation}
\sigma_0 = \frac{\left\vert q_0 \right\vert}{\sqrt{2 B}} = \frac{\left\vert q_0 \right\vert}{d(F_0,V_0;K,V_\rmin)}
\end{equation}
\begin{equation}
\frac{\sigma_1}{\sigma_0} = - \frac{\wC + \ln(\sigma_0) + \frac{1}{2} \beta_0 \ln(K F_0)}{2 B} \rlap{\ .}
\end{equation}
\begin{equation}
\frac{\sigma_2}{\sigma_0} = \frac{3}{2} \left(\frac{\sigma_1}{\sigma_0}\right)^2 - \frac{1}{2B} \left( \wD + 3\frac{\sigma_1}{\sigma_0} - \frac{\beta_0 (2-\beta_0) \sigma_0^2}{8 K^{1-\beta_0} F_0^{1-\beta_0}}\right) \rlap{\ .}
\end{equation}
This gives the CEV implied volatility expansion
\begin{equation}
\sigma = \sigma_0 \left( 1 + \frac{\sigma_1}{\sigma_0} \, T + \frac{\sigma_2}{\sigma_0} \, T^2 + o(T^2) \right) \rlap{\ .}
\end{equation}

The Black implied volatility formulas correspond to the special case $\beta_0=1$. The Bachelier (\emph{i.e.} normal) implied volatility would correspond to $\beta_0=0$.

\subsection{Generalization}

This technique can be generalized easily to other parameterizations of the options prices. Consider a model with local volatility or stochastic volatility, for which there are closed form formulas for European option prices. It can be used as a proxy in the following way.
\begin{itemize}
  \item Denoting by $z_i$ the parameters of the model, compute $B_*(z_i)$, $\wC_*(z_i)$ and $\wD_*(z_i)$, the quantities $B$, $\wC$ and $\wD$ of the asymptotic expansion~\eqref{eq-time-value} for this model at a given strike.
  \item Find parameters $z_i^{(0)}$ such that $B_*(z_i^{(0)}) = B$ (there can be several solutions).
  \item Choose a one-dimensional subset of the parameters $z_i = z_i(\lambda)$ which allows a wide range of option prices at the given strike and such that $z_i(0) = z_i^{(0)}$.
  \item Compute derivatives of $B_*(z_i)$, $\wC_*(z_i)$ and $\wD_*(z_i)$ with respect to $\lambda$ at $\lambda=0$. We use the notation $B_* = B_*(z_i(0))$, $B_*' = \partial_\lambda B_*(z_i(\lambda))\mid_{\lambda=0}$ \ldots
  \item Write a Taylor expansion $\lambda(T) = \lambda_1 T + \lambda_2 T^2 + o(T^2)$ and write the equality of the asymptotic expansion~\eqref{eq-time-value} for the model and the proxy model:
      \begin{multline*}
        \frac{B}{T} + \wC + \ln(B) + \wD T + \frac{3}{2B} T =
        \frac{B_* + \lambda_1 T B_*' + \lambda_2 T^2 B_*' + \frac{1}{2} \lambda_1^2 T^2 B_*''}{T}
        \\
        + \wC_* + \lambda_1 T \wC_*' + \ln(B_* + \lambda_1 T B_*') + \wD_* + \frac{3}{2B_*} T
        + o(T) \rlap{\ .}
      \end{multline*}
  \item This gives the Taylor expansion of $\lambda$:
  \begin{eqnarray*}
    \lambda_1 &=& \frac{\wC-\wC_* + \ln(B) - \ln(B_*)}{B_*'}
    \\
    \lambda_2 &=& \frac{\wD-\wD_* - \lambda_1 \frac{B_*'}{B_*} - \lambda_1 \wC_*'-\frac{1}{2} \lambda_1^2 B_*''}{B_*'} \rlap{\ .}
  \end{eqnarray*}
  \item Plug parameters $z_i(\lambda_1 T + \lambda_2 T^2)$ into the closed form option price of the proxy model to get an approximate price of the option in the real model.
\end{itemize}
The closer the models are, the better the approximation is. It is clear that if the proxy model is the real model itself, there are no corrections at all. This procedure consists in approximating only the differences between models at a given strike and not the option price itself. In the basic case of section~\ref{subsec:BS} where the proxy model is the Black-Scholes model, the approximation leverages on the fact that the volatility surface is more regular than the option price.

\section{SABR Model}
\label{sec:SABR}

\subsection{Model}

The SABR Model \cite{hagan2002msr} is a stochastic volatility model where the volatility is a local volatility function multiplied by a lognormal stochastic volatility:
\begin{eqnarray*}
\ud F &=& V C(F) \ud W_1
\nonumber \\
\ud V &=& \nu V \ud W_2
\end{eqnarray*}
with $ \langle \ud W_1 \ud W_2\rangle = \rho \ud t$. The initial value for $V$ is the parameter\footnote{We use the standard notation of $\alpha$ for the initial value of the volatility variable in the SABR model instead of $V_0$ as in the previous section.} $\alpha$:
\begin{equation*}
\alpha = V(0) \rlap{\ .}
\end{equation*}
$C(F)$ is a local volatility function, which is generally
\begin{equation*}
C(F) = F^\beta
\end{equation*}
$\beta$ is a number between 0 and 1 which controls the local skew. 0 corresponds to a normal process and 1 to a lognormal process. The implied volatility at time 0 and at the money is the local volatility $\alpha F_0^\beta$.

Depending on the parameters, the origin $F=0$ could be reached with finite probability in finite time. For example this happens for the CEV process (\emph{i.e.} even without stochastic volatility) for $\beta \leq \frac{1}{2}$. If $F$ models  a positive variable, a boundary condition must be imposed. The asymptotic expansion does not distinguish between different boundary conditions, as the computation is local around the geodesic path. It is valid as long as this geodesic does not reach the boundary. However the maturity validity range may be reduced for low strikes, when the probability of reflection or absorbtion at the origin modifies the probability distribution at the strike considered in a significant way.

In the following sections, we compute the asymptotic expansion for the SABR model. This short maturity expansion is valid when both $\alpha^2 T$ and $\nu^2 T$ are small enough in front of 1. If ones uses CEV implied volatility instead of lognormal implied volatility, the expansion is in $\nu^2 T$ only. Numerical experiments indicates that the approximation remains very good for $\nu^2 T < 1$.

\subsection{Order 0: metric}

In order to compute the order 0 implied volatility, the only geometric object involved is the metric. According to the dictionary of section \ref{subsec-gauge}, its inverse is the covariance matrix
\begin{equation*}
\Big( g^{ij} \Big) = \left( \begin{array}{cc}
V^2 C(F)^2 & \rho \nu V^2 C(F)
\\
\rho \nu V^2 C(F) & \nu^2 V^2
\end{array}\right) \rlap{\ .}
\end{equation*}
This matrix is first simplified by changing the variable $F$ to
\begin{equation}
q = \int_{F_0}^F \frac{\ud F}{C(F)}
\end{equation}
which for $C(F) = F^\beta$ reads for $\beta \neq 1$
\begin{equation*}
q = \frac{F^{1-\beta}-F_0^{1-\beta}}{1-\beta}
\end{equation*}
and for $\beta = 1$
\begin{equation*}
q = \ln\!\left(\frac{F}{F_0}\right) \rlap{\ .}
\end{equation*}

In addition, we rescale the time such that $\nu$ disappears of the equations while keeping the same solution of the equations (the variances which are the physical quantities are not changed):
\begin{eqnarray*}
t & \longrightarrow & \nu^2 t
\\
\alpha & \longrightarrow & \frac{\alpha}{\nu}
\\
\nu & \longrightarrow & 1
\end{eqnarray*}
At the end of the computation, the inverse transformation must be applied to the implied volatility:
\begin{equation*}
\sigma \nu \ \longleftarrow \ \sigma
\end{equation*}

The matrix in the set of variables $(q,V)$ after this rescaling is
\begin{equation*}
\Big( g^{ij} \Big) = V^2 \left( \begin{array}{cc}
1  & \rho
\\
\rho & 1
\end{array}\right) \rlap{\ .}
\end{equation*}
This is diagonalized by going from variables $(q,V)$ to $(x,y)$ with
\begin{eqnarray*}
x &=& \frac{q-\rho V}{\sqrt{1-\rho^2}}
\\
y &=& V   \rlap{\ .}
\end{eqnarray*}
The covariance matrix becomes
\begin{equation*}
\Big( g^{ij} \Big)  = y^2 \left( \begin{array}{cc}
1  & 0
\\
0 & 1
\end{array}\right)
\end{equation*}
and its inverse is the metric
\begin{equation*}
\Big( g_{ij} \Big)  = \frac{1}{y^2} \left( \begin{array}{cc}
1  & 0
\\
0 & 1
\end{array}\right)
\end{equation*}
which corresponds to the infinitesimal distance
\begin{equation*}
\ud s^2 = \frac{\ud x^2 + \ud y^2}{y^2} \rlap{\ .}
\end{equation*}
This geometry corresponds to the hyperbolic plane, in the Poincaré half-plane representation ($y>0$) \cite{hagan2001pds,henrylabordere:gai}.
Geodesics are vertical lines and semi-circles orthogonal to the $y=0$ axis.
The geodesic distance between two points $(x_1,y_1)$ and $(x_2,y_2)$ can be computed:
\begin{equation*}
d(x_1,y_1;x_2,y_2) = \acosh\!\left( 1+ \frac{(x_2-x_1)^2+(y_2-y_1)^2}{2 y_1 y_2} \right) \rlap{\ .}
\end{equation*}
In the $(q,V)$ variables, going from $q=0,V=\alpha$ to $q,V$ the geodesic distance is
\begin{equation*}
d(0,\alpha;q,V) = \acosh\!\left( 1 + \frac{q^2+ (V-\alpha)^2 - 2\rho q (V-\alpha)}{2 (1-\rho^2) \alpha V} \right) \rlap{\ .}
\end{equation*}
For a given strike, \emph{i.e.} a given $q$, it is minimized by the volatility
\begin{equation*}
V_\rmin = \sqrt{\alpha^2 + 2 \rho \alpha q + q^2}
\end{equation*}
and the minimal distance is
\begin{equation*}
d(0,\alpha;q) = \acosh\!\left( \frac{V_\rmin - \rho q - \rho^2 \alpha}{(1-\rho^2) \alpha}\right) = \left\vert \ln\!\left(\frac{V_\rmin + \rho \alpha + q}{(1+\rho) \alpha}\right) \right\vert \rlap{\ .}
\end{equation*}

Equation \eqref{eq-order0} gives the order 0 implied volatility
\begin{equation*}
\sigma_0 = \frac{\displaystyle \ln\!\left(\frac{K}{F_0}\right)}{\displaystyle \ln\!\left(\frac{V_\rmin + \rho \alpha + q}{(1+\rho) \alpha}\right)} \rlap{\ .}
\end{equation*}
(We have dropped the absolute values as the numerator and the denominator have the same sign.)

Plugging the expression for $V_\rmin$ and going back to the original time, with $\nu$ factors, the order 0 implied volatility for the SABR model is
\begin{equation}
\sigma_0 = \frac{\displaystyle \nu \ \ln\!\left(\frac{K}{F_0}\right)}{\displaystyle \ln\!\left(\frac{\sqrt{\alpha^2 + 2 \rho \alpha \nu q + \nu^2 q^2} + \rho \alpha + q \nu}{(1+\rho) \alpha}\right)}
\label{eq-SABR-0}
\end{equation}
with
\begin{equation*}
q = \left\{ \begin{array}{ll}
\displaystyle \frac{K^{1-\beta}-F_0^{1-\beta}}{1-\beta} & \beta < 1
\\
\displaystyle \ln\!\left(\frac{K}{F_0}\right) & \beta = 1
\rlap{\ .}
\end{array}
\right.
\end{equation*}

At the money, the limit of this expression is simply
\begin{equation*}
\sigma_0(F_0) = \alpha F_0^{\beta-1}
\end{equation*}
which is the local volatility.

\subsection{Order 1: connection}
\label{subsec:order1}

To compute the order 1 correction, we need the scalar factor in the time value expansion, given by $\wC$ in equation \eqref{eq-expected-C}, with $C$ given in equation \eqref{eq-saddle-integrand-C}.

For the hyperbolic plane, the Van Vleck--Morette determinant can be computed as a function of the geodesic distance:
\begin{equation*}
\Delta = \frac{d}{\sinh(d)} \rlap{\ .}
\end{equation*}
We need also
\begin{eqnarray*}
\sigma_F(K,V) &=& V C(K) = V K^\beta
\\
g(K,V) &=& \frac{1}{V^4 C(K)^2 (1-\rho^2)} = \frac{1}{V^4 K^{2\beta} (1-\rho^2)}
\end{eqnarray*}
Using that $d'=0$ for the minimum distance at $V=V_\rmin$, we compute
\begin{equation*}
B'' = d d'' = \frac{d}{\alpha V_\rmin (1-\rho^2) \sinh(d) }
\end{equation*}
Terms simplify against each other to give at $V=V_\rmin$
\begin{equation}
\wC = -\ln\!\left(\sqrt{\alpha  V_\rmin} K^\beta \right) + \cA
\label{eq-wC-interm}
\end{equation}

$\cA$ is the integral of the connection 1-form on the geodesic. According to formula \eqref{eq-A}, the connection is given by
\begin{equation}
A = \frac{1}{2(1-\rho^2)} \, \ud \ln(C(F)) - \frac{\rho C'(F)}{2(1-\rho^2)} \, \ud V 
\rlap{\ .}
\label{eq-A-integr}
\end{equation}
In fact, $A$ can be rewritten in $(x,y)$ variables as
\begin{equation*}
A = \frac{C'(F)}{2 \sqrt{1-\rho^2}} \, \ud x
\rlap{\ .}
\end{equation*}
It must be integrated from
\begin{equation*}
\left( \begin{array}{c}
x_1 \\ y_1
\end{array} \right)
=
\left( \begin{array}{c}
\frac{-\rho \alpha}{\sqrt{1-\rho^2}} \\ \alpha
\end{array} \right)
\end{equation*}
to
\begin{equation*}
\left( \begin{array}{c}
x_2 \\ y_2
\end{array} \right)
=
\left( \begin{array}{c}
\frac{q-\rho V}{\sqrt{1-\rho^2}} \\ V
\end{array} \right)
\rlap{\ .}
\end{equation*}

If $\beta=1$, the 1-form
\begin{equation*}
A = \frac{1}{2 \sqrt{1-\rho^2}} \, \ud x
\end{equation*}
is exact and can be integrated directly:
\begin{equation*}
\cA = \frac{q-\rho V + \rho \alpha}{2 \left(1-\rho^2\right)} = \frac{1}{2} \ln\!\left(\frac{K}{F_0}\right) + \frac{\rho}{2 \left(1-\rho^2\right)} \left( \rho \ln\!\left(\frac{K}{F_0}\right) - V + \alpha \right) \rlap{\ .}
\end{equation*}

We consider now the general case where $\beta <1$. If $x_1=x_2$, the geodesic is a vertical line. As $A$ is along $\ud x$, its integral is zero:
\begin{equation*}
\cA = 0 \rlap{\ .}
\end{equation*}
In other cases, the first part of equation \eqref{eq-A-integr} is an exact form and can be integrated directly:
\begin{equation}
\int_{(F_0,\alpha)}^{(K,V)} \frac{1}{2(1-\rho^2)} \, \ud \ln(C(F)) = \frac{1}{2(1-\rho^2)} \, \ln\!\left( \frac{C(K)}{C(F)}\right)
= \frac{\beta}{2(1-\rho^2)} \, \ln\!\left( \frac{K}{F}\right)
\rlap{\ .}
\label{eq-cA-1}
\end{equation}
The second part must be integrated on the geodesic path.
The geodesic is a semi-circle with origin $(X,0)$, radius $R$ and going through $(x_1,x_2)$ and $(y_1,y_2)$.
The origin is therefore
\begin{equation}
X = \frac{x_2^2-x_1^2 + y_2^2-y_1^2}{2(x_2-x_1)}
\label{eq-X}
\end{equation}
and the radius
\begin{equation}
R = \sqrt{y_1^2 + (x_1-X)^2} \rlap{\ .}
\label{eq-R}
\end{equation}
We parameterize the geodesic by $t=\tan(\theta/2)$ where $\theta$ is the angle on the circle:
\begin{eqnarray*}
x &=& X + R \, \frac{1-t^2}{1+t^2}
\\
y &=& R \, \frac{2t}{1+t^2} \rlap{\ .}
\end{eqnarray*}
In this parametrization, the geodesic distance is given by
\begin{equation*}
\ud s = \ud \ln(t)
\end{equation*}
and we can compute
\begin{equation}
- \int_\cC \frac{\rho \beta F^{\beta-1}}{2(1-\rho^2)} \ud V = - \frac{\rho^2 \beta}{2(1-\rho^2)} \, \ln\!\left( \frac{K}{F}\right)
- \frac{\rho \beta}{(1-\beta)\sqrt{1-\rho^2}} \left[ G(t_2)-G(t_1) \right]
\label{eq-cA-2}
\end{equation}
with
\begin{multline}
G(t) = \atan(t)
\\
+ \left\{
\begin{array}{ll}
\displaystyle - \frac{a+bX}{\sqrt{(a+bX)^2-(1-\beta)^2 R^2}} \atan\!\left(\frac{cR+t(a+b(X-R))}{\sqrt{(a+bX)^2-(1-\beta)^2 R^2}} \right) & (a+bX)^2 > (1-\beta)^2 R^2
\\
\displaystyle \frac{a+b X}{c R+t(a+b(X-R))} & (a+bX)^2 = (1-\beta)^2 R^2
\\
\displaystyle \frac{a+bX}{\sqrt{(1-\beta)^2 R^2-(a+bX)^2}} \, \widetilde{\atanh}\!\left(\frac{cR+t(a+b(X-R))}{\sqrt{(1-\beta)^2 R^2-(a+bX)^2}} \right) & (a+bX)^2 < (1-\beta)^2 R^2
\end{array}
\right.
\label{eq-G}
\end{multline}
\begin{eqnarray*}
a &=& F_0^{1-\beta}
\\
b &=& (1-\beta) \sqrt{1-\rho^2}
\\
c &=& (1-\beta) \rho
\end{eqnarray*}
\begin{equation}
t_i = \sqrt{\frac{R-x_i+X}{R+x_i-X}}
\label{eq-t}
\end{equation}
and
\begin{equation*}
\widetilde{\atanh}(z) = \frac{1}{2} \ln\!\left\vert \frac{1+z}{1-z} \right\vert \rlap{\ ,}
\end{equation*}
which coincides with the inverse function of $\tanh$ on $] -1 ; 1[$.
Summing equations \eqref{eq-cA-1} and \eqref{eq-cA-2}, the integral of the connection is finally
\begin{equation}
\cA = \frac{\beta}{2} \, \ln\!\left( \frac{K}{F}\right)
- \frac{\rho \beta}{(1-\beta)\sqrt{1-\rho^2}} \left[ G(t_2)-G(t_1) \right]
\rlap{\ .}
\label{eq-cA}
\end{equation}

Replacing $\cA$ in equation \eqref{eq-wC-interm} we get
\begin{equation}
\wC = -\frac{1}{2}\ln\!\left(\alpha  F_0^\beta \, V_\rmin K^\beta \right) + \left\{ \begin{array}{ll}
\displaystyle - \frac{\rho \beta}{(1-\beta)\sqrt{1-\rho^2}} \left[ G(t_2)-G(t_1) \right] & \beta < 1
\\
\displaystyle \frac{\rho}{2 \left(1-\rho^2\right)} \left( \rho \ln\!\left(\frac{K}{F_0}\right) - V_\rmin + \alpha \right) & \beta=1
\end{array} \right.
\label{eq-wC}
\end{equation}
Restoring the factor $\nu$, the order 1 correction is given by equation \eqref{eq-order1}:
\begin{equation}
\frac{\sigma_1}{\sigma_0} = - \nu^2 \frac{\displaystyle \wC + \ln\!\left(\frac{\sigma_0}{\nu} \sqrt{K F_0}\right)}{\displaystyle \ln^2\!\left(\frac{\sqrt{\alpha^2 + 2 \rho \alpha \nu q + \nu^2 q^2} + \rho \alpha + q \nu}{(1+\rho) \alpha}\right)}
\label{eq-SABR-1}
\end{equation}
$\sigma_0$ is given by equation \eqref{eq-SABR-0}, $\wC$ by equation \eqref{eq-wC} (with $\alpha$ divided by $\nu$, also inside $V_\rmin$), where $X$, $R$, $t_1$ and $t_2$ are given in equations \eqref{eq-X}, \eqref{eq-R} and \eqref{eq-t} from
\begin{equation*}
\left( \begin{array}{c}
x_1 \\ y_1
\end{array} \right)
=
\left( \begin{array}{c}
\displaystyle \frac{-\rho \alpha}{\nu \sqrt{1-\rho^2}} \\ \displaystyle \frac{\alpha}{\nu}
\end{array} \right)
\textrm{\ and \ }
\left( \begin{array}{c}
x_2 \\ y_2
\end{array} \right)
=
\left( \begin{array}{c}
\displaystyle \frac{\nu q-\rho \sqrt{\alpha^2 + 2 \rho \alpha \nu q + \nu^2 q^2}}{\nu \sqrt{1-\rho^2}} \\ \displaystyle \frac{\sqrt{\alpha^2 + 2 \rho \alpha \nu q + \nu^2 q^2}}{\nu}
\end{array} \right)
\end{equation*}
and $G(t)$ is defined in formula \eqref{eq-G}.

Using this expression, the first order implied volatility is
\begin{equation*}
\sigma = \sigma_0 \left( 1 + \frac{\sigma_1}{\sigma_0} t + o(t) \right)
\end{equation*}
which is valid for all positive strikes.
Exactly at the money, the formula we give must be replaced by its limit, which can be computed by a Taylor expansion or numerically. At the money and only at the money it appears to be equal to the original HKLW formula:
\begin{equation}
\frac{\sigma_1}{\sigma_0}(F_0) = \frac{1}{24} \alpha^2 (1-\beta)^2 F_0^{2(\beta-1)} + \frac{1}{4} \rho \alpha \nu \beta F_0^{\beta-1} + \frac{1}{12} \nu^2 - \frac{1}{8} \rho^2 \nu^2 \rlap{\ .}
\label{eq:order1atm}
\end{equation}
This is not surprising as their expansion is in fact an expansion in both maturity and moneyness (eventually of order 0 in moneyness).

\subsection{Order 2}

To compute the second order correction to implied volatility, we need to compute $\wD$ as defined in equation \eqref{eq-expected-D}, with $D=-a_1$ defined in equation \eqref{eq-a1}.

We have to compute $a_1$ as defined in equation \eqref{eq-a1}. Most of the integration can be done analytically.
We have first the integral of $Q$ along the geodesic:
\begin{equation*}
a_1^{(Q)} = - \frac{1}{d} \int_\cC Q \ud s \rlap{\ .}
\end{equation*}
According to equation \eqref{eq-Q}, $Q$ is
\begin{equation*}
Q = \frac{\beta}{4} \left( 1-\beta + \frac{\beta}{2(1-\rho^2)} \right) \frac{V^2}{F^{2(1-\beta)}} \rlap{\ .}
\end{equation*}
Using the values defined in the previous section for $X$, $R$, $t_1$, $t_2$, $a$, $b$ and $c$, its integral along the geodesic is
\begin{equation}
a_1^{(Q)} = \frac{\beta}{2} \left(1-\beta + \frac{\beta}{2(1-\rho^2)} \right)\frac{R^2}{(1-\beta)^2 R^2-(a+bX)^2} \, \frac{ H(t_2) - H(t_1) }{\ln(t_2)-\ln(t_1)}
\label{eq-a1-Q}
\end{equation}
with
\begin{multline*}
H(t) = \frac{a + b (R+X) + c R t}{(a +b X) (1+t^2) + b R (1-t^2) + 2 c R t}
\\
+
\left\{
\begin{array}{ll}
\displaystyle \frac{c R}{\sqrt{(a+bX)^2-(1-\beta)^2 R^2}} \atan\!\left(\frac{cR+t(a+b(X-R))}{\sqrt{(a+bX)^2-(1-\beta)^2 R^2}} \right) & (a+bX)^2 > (1-\beta)^2 R^2
\\
\displaystyle -\frac{c R}{\sqrt{(1-\beta)^2 R^2-(a+bX)^2}} \, \widetilde{\atanh}\!\left(\frac{cR+t(a+b(X-R))}{\sqrt{(1-\beta)^2 R^2-(a+bX)^2}} \right) & (a+bX)^2 < (1-\beta)^2 R^2
\ .
\end{array}
\right.
\end{multline*}
Note that in the denominator, the quantity $\ln(t_2)-\ln(t_1)$ is up to a sign the geodesic distance $d$.
If $\beta=1$, $a_1^{(Q)}$ reduces to
\begin{equation}
a_1^{(Q)} = - \frac{R \mid x_2-x_1 \mid }{8(1-\rho^2)d} \rlap{\ .}
\label{eq-a1-Q-bis}
\end{equation}

The Laplacian on the hyperbolic plane is in $(x,y)$ coordinates
\begin{equation*}
D^i D_i = y^2 (\partial_x^2 + \partial_y^2) \rlap{\ .}
\end{equation*}
As the Van Vleck--Morette determinant $\Delta = \frac{s}{\sinh(s)}$ depends only on the geodesic distance $s$, its derivative on the orthogonal coordinate vanishes: $\partial_\perp \Delta = 0$. On the other hand, by definition the parallel transport on the geodesic curve has no covariant derivative along the curve: $\nabla_s \cP = 0$. As a consequence, there is no crossed term and both terms decouple: we have
\begin{equation*}
\cP^{-1} \Delta^{-1/2} \nabla^i \nabla_i ( \cP \Delta^{1/2} ) = \cP^{-1} \nabla^i \nabla_i \cP + \Delta^{-1/2} D^i D_i \Delta^{1/2}
\end{equation*}
(the $\RR$ charge is carried only by $\cP$).

The metric part can be integrated analytically \cite{hagan2001pds,avramidi:agm}:
\begin{equation}
a_1^{(R)} = -\frac{1}{8}\left[ 1 + \frac{1}{d} \left( \frac{\cosh(d)}{\sinh(d)} - \frac{1}{d}\right) \right] \rlap{\ .}
\label{eq-a1-R}
\end{equation}

The last part to integrate is the $\RR$ connection term $\cP^{-1} \nabla^i \nabla_i \cP$. As the action of gauge transformations on the heat kernel expansion is fully carried by the parallel transport term $\cP$, $a_1$ can only depend on gauge-invariant quantities constructed from $F=\ud A$. We split therefore $A = A^{(0)} + A^{(1)}$ into a pure gauge part $A^{(0)}$ and $A^{(1)}$ such that $F = \ud A^{(1)}$:
\begin{eqnarray*}
A^{(0)} &=& \frac{1}{2} \ud \ln(C(F))
\\
A^{(1)} &=& \frac{\rho^2}{2(1-\rho^2)} \, \ud \ln(C(F)) - \frac{\rho C'(F)}{2(1-\rho^2)} \, \ud V \rlap{\ .}
\end{eqnarray*}
Forgetting $A^{(0)}$ which is pure gauge (it can be checked by hand that $A^{(0)}$ will not contribute), we denote by $\cP^{(1)}$ the $A^{(1)}$ part of the parallel transport:
\begin{equation*}
\cP^{(1)} = e^{-\cA^{(1)}} = e^{-\int_\cC A^{(1)} }
\end{equation*}
with
\begin{equation}
\cA^{(1)} = \left\{ \begin{array}{ll}
\displaystyle - \frac{\rho \beta}{(1-\beta)\sqrt{1-\rho^2}} \left[ G(t_2)-G(t_1) \right] & \beta < 1
\\
\displaystyle \frac{\rho}{2 \left(1-\rho^2\right)} \left( \rho \ln\!\left(\frac{K}{F_0}\right) - V + \alpha \right) & \beta=1
\rlap{\ .}
\end{array}
\right.
\label{eq-cA1}
\end{equation}
Using $\cA^{(1)}$ and $A^{(1)}$ which in $(x,y)$ variables is
\begin{equation*}
A^{(1)} = \frac{\rho C'(F)}{2} \left( \frac{\rho}{\sqrt{1-\rho^2}} \, \ud x - \ud y \right) \rlap{\ ,}
\end{equation*}
we can rewrite
\begin{equation*}
\frac{1}{2} \cP^{-1} \nabla^i \nabla_i \cP = \frac{1}{2} y^2 \left[ - \left(\partial_x^2+\partial_y^2 \right)\cA^{(1)} + \left( \partial_x \cA^{(1)} - A^{(1)}_x \right)^2 + \left( \partial_y \cA^{(1)} - A^{(1)}_y \right)^2 + \partial_x A^{(1)}_x + \partial_y A^{(1)}_y \right]
\end{equation*}
where the specific form of $A^{(1)}$ can be used to simplify terms
\begin{equation*}
\partial_x A^{(1)}_x + \partial_y A^{(1)}_y = 0 \rlap{\ .}
\end{equation*}

Computing $\cA^{(1)}$ and $A^{(1)}$ analytically, we compute numerically the $x$ and $y$ derivatives and integrate numerically along the geodesic curve to get
\begin{equation}
a_1^{(A)} = \frac{1}{2 d} \int_\cC  \ud s \, y^2 \left[ - \left(\partial_x^2+\partial_y^2 \right)\cA^{(1)} + \left( \partial_x \cA^{(1)} - A^{(1)}_x \right)^2 + \left( \partial_y \cA^{(1)} - A^{(1)}_y \right)^2 \right] \rlap{\ .}
\label{eq-a1-A}
\end{equation}
This integral can be computed by a numerical quadrature with few points. For $\beta=1$ the connection has in fact no curvature and therefore $a_1^{(A)} =0$.

We compute also the following quantities (at $V=V_\rmin$):
\begin{eqnarray}
d'' &=& \frac{1}{\alpha (1-\rho^2) V_\rmin \sinh(d)}
\nonumber\\
B'' &=& d d''
\nonumber\\
\frac{B^{(3)}}{B''} &=& - \frac{3}{V_\rmin}
\label{eq-B2}
\\
\frac{B^{(4)}}{B''} &=& \frac{12}{V_\rmin^2} - 3 d'' \left( \frac{\cosh(d)}{\sinh(d)} - \frac{1}{d}\right) \rlap{\ .}
\nonumber
\end{eqnarray}
We need finally $\wC'$ and $\wC''$.
Using our decomposition $\cA = \frac{\beta}{2} \ln \frac{K}{F} + \cA^{(1)}$, we can write from formulas \eqref{eq-saddle-integrand-C} and \eqref{eq-expected-C}
\begin{equation*}
\wC = - \frac{\beta}{2} \ln(KF) - \frac{1}{2} \ln\!\left(\frac{d}{\sinh(d)}\right) + \frac{1}{2}\ln\!\left(B''\right) + \ln\!\left(1-\rho^2\right) + \cA_1 \rlap{\ .}
\end{equation*}
Its first and second derivatives in $V$ at $V=V_\rmin$ are
\begin{eqnarray*}
\wC' &=& \frac{1}{2} \frac{B^{(3)}}{B''} + {\cA^{(1)}}'
\\
\wC'' &=& \frac{1}{2} d'' \left( \frac{\cosh(d)}{\sinh(d)} - \frac{1}{d}\right) + \frac{1}{2} \frac{B^{(4)}}{B''} - \frac{1}{2} \left( \frac{B^{(3)}}{B''}\right)^2 + {\cA^{(1)}}''
\rlap{\ .}
\end{eqnarray*}
We choose to differentiate $\cA^{(1)}$ numerically, by finite difference, as the analytical expression is very long and hard to simplify.

Simplifying $\frac{\wC''}{2B''}$ against $-a_1^{(R)}$ we get finally
\begin{equation*}
\wD = - a_1^{(Q)} - a_1^{(A)} + \frac{1}{8} + \frac{1}{2 B''}\left[ {\cA^{(1)}}'' - {\cA^{(1)}}'^2 + \frac{3}{V_\rmin} {\cA^{(1)}}' - \frac{3}{4 V_\rmin^2} \right]
\label{eq-wD}
\end{equation*}
with $a_1^{(Q)}$ given in equation \eqref{eq-a1-Q}, $a_1^{(A)}$ in \eqref{eq-a1-A}, $B''$ in \eqref{eq-B2} and $\cA^{(1)}$ in \eqref{eq-cA1} with $G(t)$ defined in equation \eqref{eq-G}. For $\beta=1$, this expression can be simplified using equation \eqref{eq-a1-Q-bis} for $a_1^{(Q)}$ and
\begin{eqnarray*}
a_1^{(A)} &=& 0
\\
{\cA^{(1)}}' &=& -\frac{\rho}{2(1-\rho^2)}
\\
{\cA^{(1)}}'' &=& 0 \rlap{\ .}
\end{eqnarray*}

We have finally obtained the second order corrective term
\begin{equation*}
\frac{\sigma_2}{\sigma_0} = \frac{3}{2} \left(\frac{\sigma_1}{\sigma_0}\right)^2 - \frac{1}{2B} \left( \wD + 3\frac{\sigma_1}{\sigma_0} - \frac{\sigma_0^2}{8}\right) \end{equation*}
such that we get the quadratic approximation to implied volatility
\begin{equation*}
\sigma = \sigma_0 \left( 1 + \frac{\sigma_1}{\sigma_0} T + \frac{\sigma_2}{\sigma_0} T^2 + o(T^2) \right) \rlap{\ .}
\end{equation*}

The computation has been done in redefined variables such that $\nu=1$. To restore the $\nu$ factors, $\alpha$ must be replaced by $\frac{\alpha}{\nu}$, $T$ by $\nu^2 T$ and the final implied volatility must be multiplied by $\nu$.

At the money, the formula for $\frac{\sigma_2}{\sigma_0}$ looks divergent but its limit is well defined. We compute this limit numerically, although it could be done analytically.

\subsection{CEV volatility}

The results of section~\ref{subsec:cev-theo} can be used to invert the SABR volatility into a CEV fractional volatility. Using formulas of section~\ref{subsec:cev-theo} the implied CEV volatility is computed and used in the closed-form option prices of the CEV model.

This appears to be useful at low strikes for $\beta<1/2$ or with small volatility of volatility: only the corrections to the CEV model which come from the stochastic volatility are approximated, not the local volatility part. For example, at the money the first order coefficient of the Black volatility which is given by equation~\eqref{eq:order1atm} becomes for the CEV volatility
\begin{equation*}
\frac{\sigma_1}{\sigma_0}(F_0) = \frac{1}{4} \rho \alpha \nu \beta F_0^{\beta-1} + \frac{1}{12} \nu^2 - \frac{1}{8} \rho^2 \nu^2 \rlap{\ .}
\end{equation*}
The corrective term in $\alpha^2 T$ has disappeared.

\subsection{Numerical results}

\begin{figure}[p]
\centering
2.5 years\\
\includegraphics[width=0.49\textwidth]{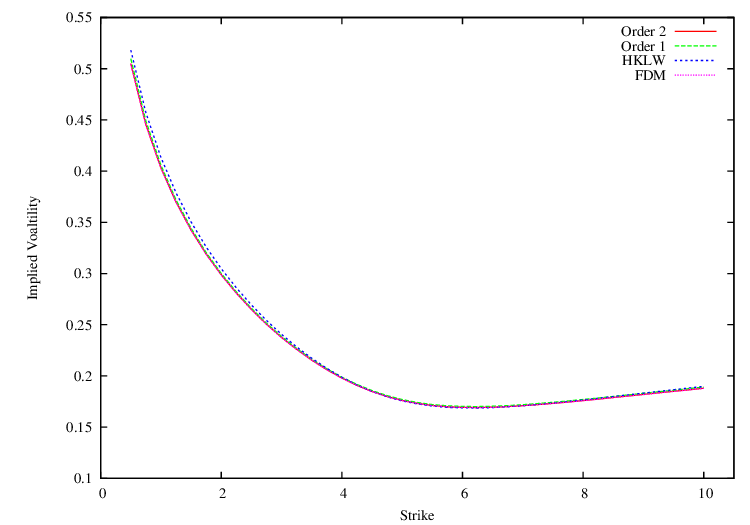}
\includegraphics[width=0.49\textwidth]{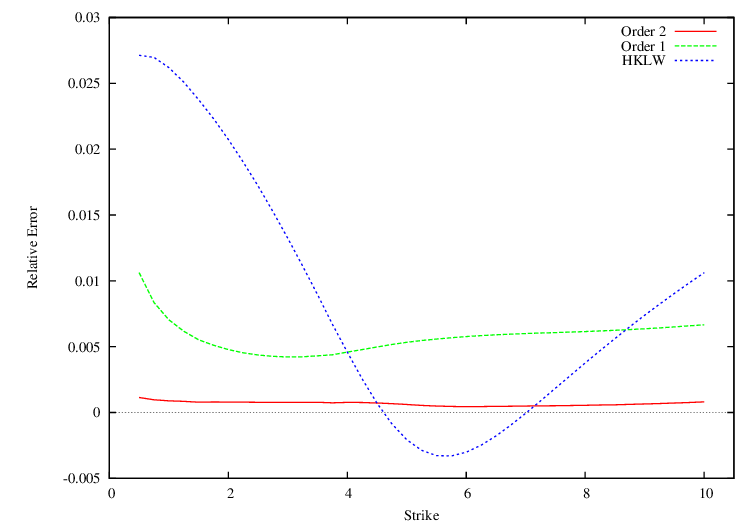}

5 years\\
\includegraphics[width=0.49\textwidth]{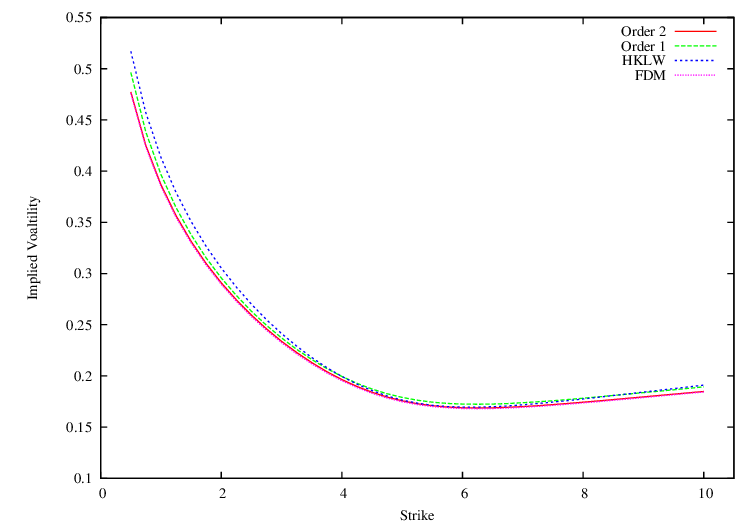}
\includegraphics[width=0.49\textwidth]{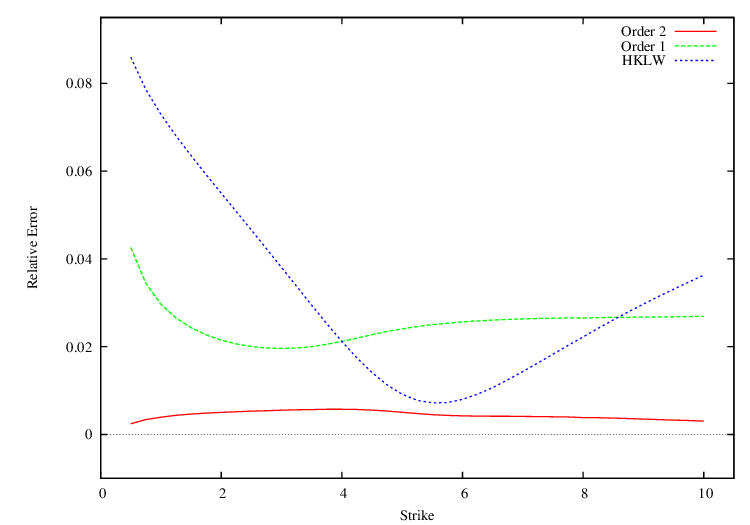}

7.5 years\\
\includegraphics[width=0.49\textwidth]{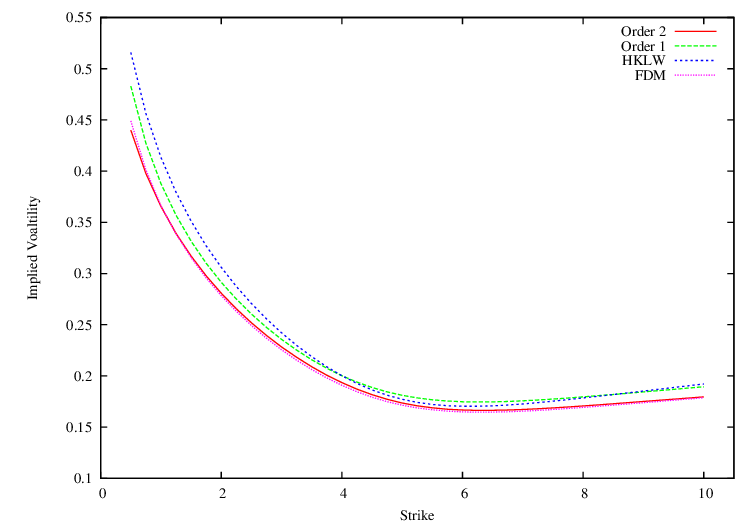}
\includegraphics[width=0.49\textwidth]{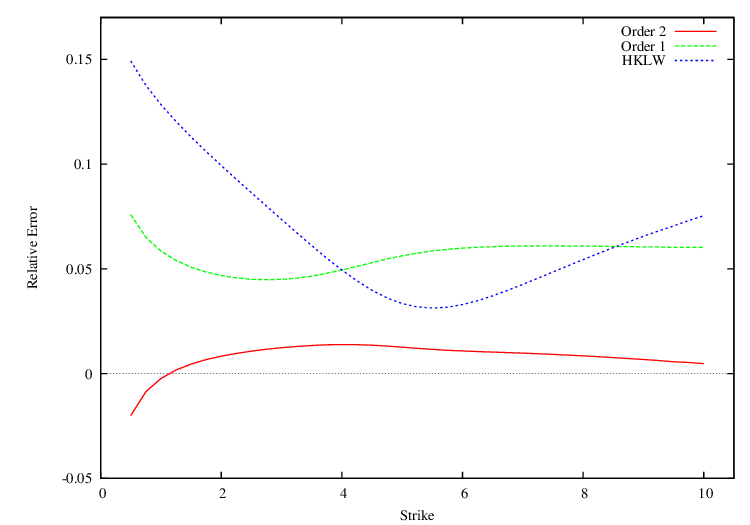}

10 years\\
\includegraphics[width=0.49\textwidth]{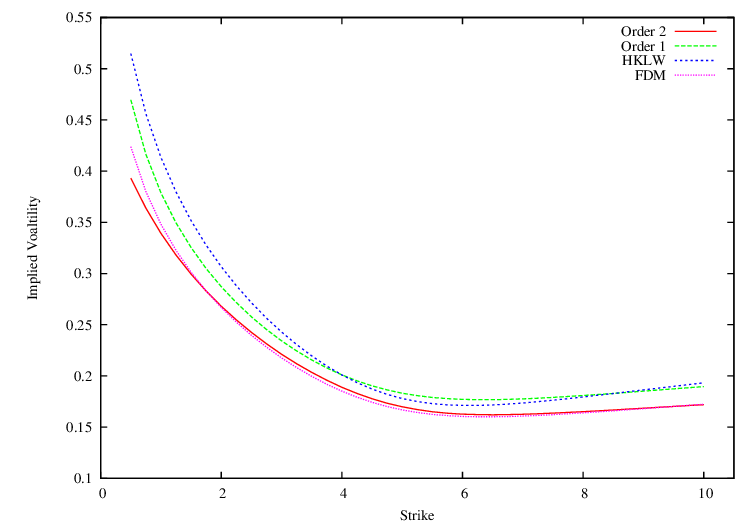}
\includegraphics[width=0.49\textwidth]{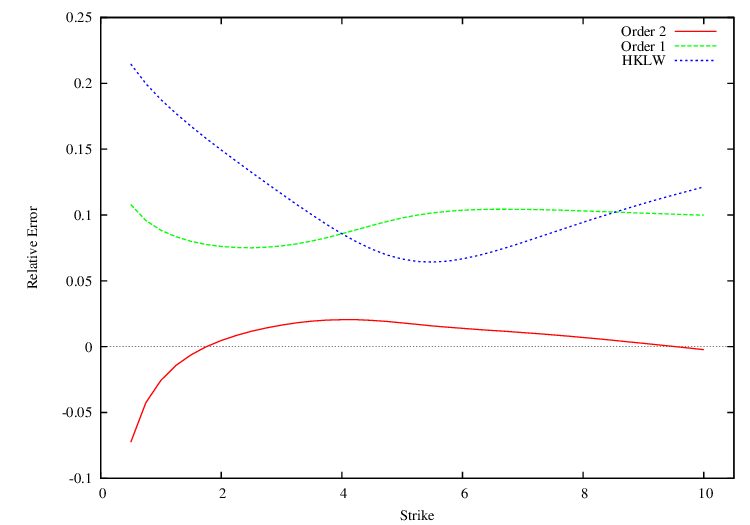}

\caption{\emph{Implied volatility and relative error for the SABR model with parameters $F_0=4$, $\alpha=30\%$, $\beta=.7$, $\nu=40\%$, $\rho=-.5$ and maturities 2.5~yr, 5~yr, 7.5~yr and 10~yr. On the left, implied volatilities are plotted for our first order and second order expansions, the original formula of \cite{hagan2002msr} and are compared to the result of a FDM solution. On the right are the relative errors with respect to this reference solution.}}
\label{fig:num1}
\end{figure}

\begin{figure}[p]
\centering
15 years\\
\includegraphics[width=0.49\textwidth]{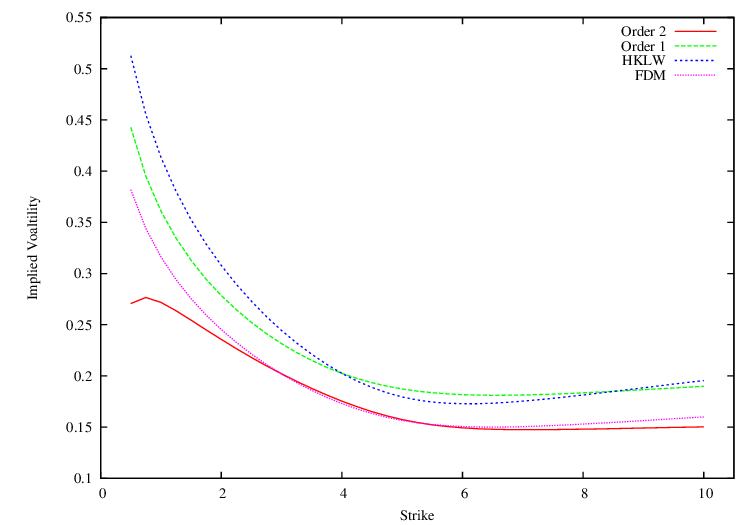}
\includegraphics[width=0.49\textwidth]{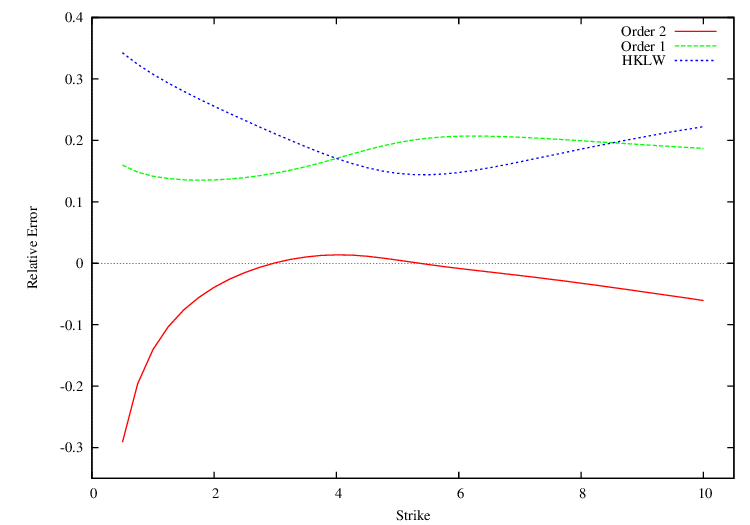}

20 years\\
\includegraphics[width=0.49\textwidth]{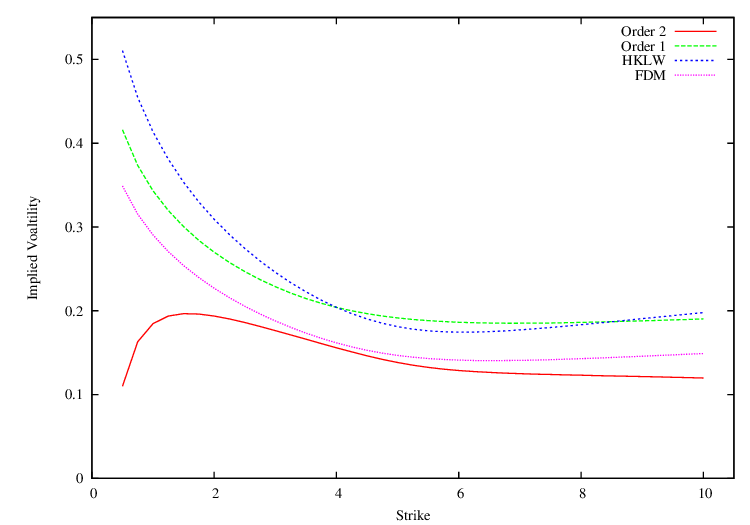}
\includegraphics[width=0.49\textwidth]{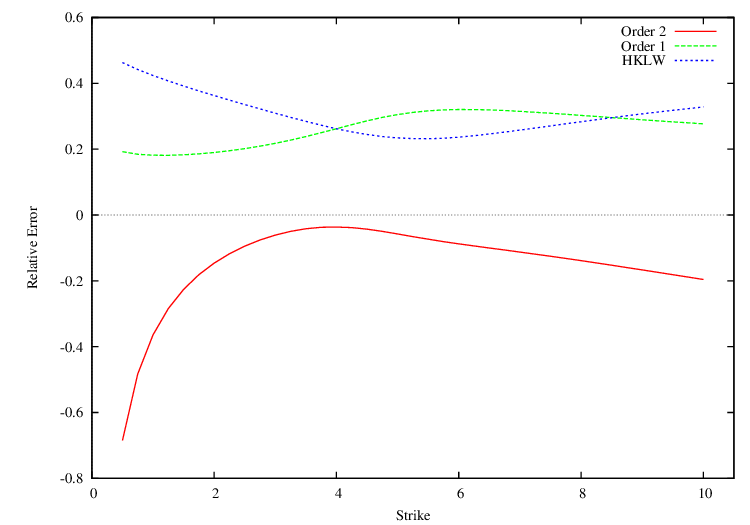}

30 years\\
\includegraphics[width=0.49\textwidth]{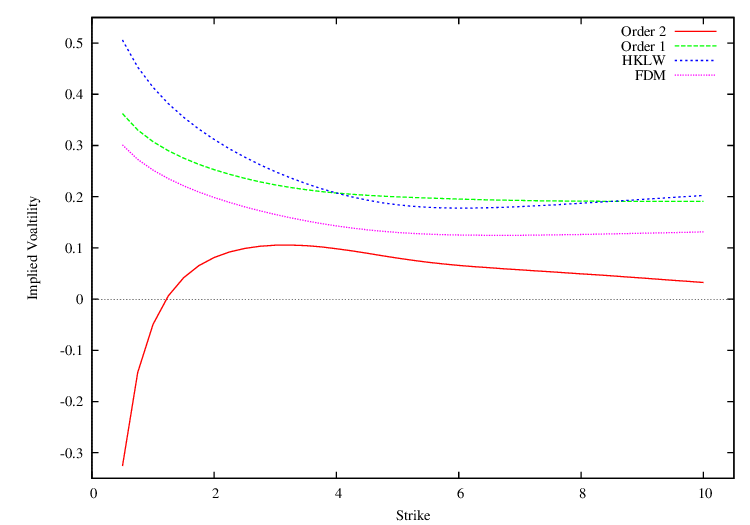}
\includegraphics[width=0.49\textwidth]{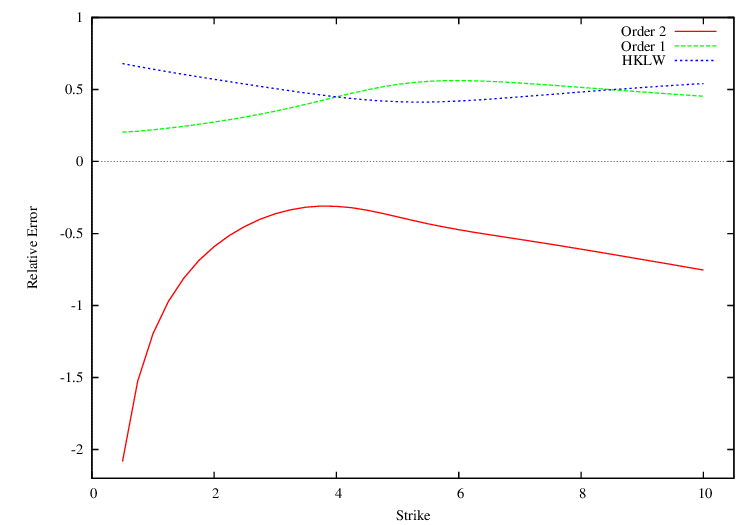}

\caption{\emph{Implied volatility and relative error for the SABR model with parameters $F_0=4$, $\alpha=30\%$, $\beta=.7$, $\nu=40\%$, $\rho=-.5$ and maturities 10~yr, 20~yr and 30~yr. On the left, implied volatilities are plotted for our first order and second order expansions, the original formula of \cite{hagan2002msr} and are compared to the result of a FDM solution. On the right are the relative errors with respect to this reference solution.}}
\label{fig:num2}
\end{figure}

We present in figures~\ref{fig:num1} and~\ref{fig:num2} the implied volatility given by our expansion and compare it to the implied volatility computed by a two-dimensional finite difference method scheme. We also show for comparison the implied volatility given by the original formula of \cite{hagan2002msr}. In this example, parameters are $F_0=4$, $\alpha=30\%$, $\beta=.7$, $\nu=40\%$, $\rho=-.5$ The FDM scheme is a second order Yanenko scheme \cite{yanenko1971method} with exponential fitting. We use 400 points in strike, 200 points in volatility and 30 time steps. 

At very short maturities, all expansions are acceptable as the expansion is dominated by the order 0 term. At first order our expansion is equal to the HKLW at the money but is more regular in strikes and is better in the wings as our computation does not involve any approximation in the moneyness. Our second order expansion is one order of magnitude more precise\footnote{In fact at very short maturities, the FDM scheme we use is less precise and less stable than this second order expansion, especially in the wings where the probability density is very small.}. When maturity grows, first order expansions lose precision but the second order remain relatively good up to 10 years, where $\nu^2 T = 1.6$. At higher maturities, the second order expansion explodes quadratically and finally gives even negative volatilities at very long maturity and low strikes. At long maturities, a FDM or an other numerical method must be used, unless a valid long maturity expansion could be computed more efficiently.

\section{Conclusion}

We have presented a general method to compute a Taylor expansion in maturity of implied volatility for stochastic volatility models. We give exact formulas for the first and second order corrections. As an application, we have computed this expansion for the SABR model and compared it to the implied volatility given by a numerical scheme and to the original HKLW formula. It appears that it gives more precise results than the usual formula and extends the domain where a short maturity expansion can be used. Outside this range of validity, other methods must be used: numerical schemes or possibly other approximations.

If a closer model with closed formulas than Black-Scholes exists, we provide a method to use this model as a proxy to extend the domain of validity of the expansion. It would be interesting to see the results of this method for the SABR model with a stochastic volatility model as a proxy.

Obtaining exact option prices at all maturities would be a non-perturbative computation, which is a longstanding issue in theoretical physics.

\FloatBarrier

\section*{Acknowledgements}

We thank Erwan Curien for his interest in this work, Martial Millet, Xavier Lacroze and Wafaa Bennehhou for many discussions.

\begin{appendices}
\section{Mean reversion}

At long maturity, the SABR model is not realistic as the volatility process is a geometric Brownian motion. In particular the variance of the volatility increases linearly in time. A direct extension would be to add mean reversion to the volatility process, either on the volatility or on the variance. The asymptotic expansion can still be computed at order 2. However its domain of validity is usually reduced: in addition to other conditions, the maturity must be small compared to the mean reversion characteristic time.

We impose mean reversion on the volatility process as
\begin{equation*}
  \ud V = \nu V \ud W_2 + \kappa(\overline V - V) \ud t
\end{equation*}
The metric is not modified as it describes the diffusion part. Expressions for $A$ and $Q$ are modified as follows:
\begin{equation*}
  A = A_{\mathrm{SABR}} - \frac{\rho \kappa(V - \overline V)}{\nu V^2 (1-\rho^2) F^\beta} \ud F + \frac{\kappa(V - \overline V)}{\nu^2 V^2 (1-\rho^2) } \ud V
\end{equation*}
\begin{equation*}
  Q = Q_{\mathrm{SABR}} + \frac{1}{2} \frac{\kappa^2(V - \overline V)^2}{\nu^2 V^2 (1-\rho^2)} + \frac{1}{2} \kappa - \kappa \frac{\overline V}{V} - \frac{1}{2} \frac{\rho \beta \kappa (V - \overline V)}{\nu (1-\rho^2) F^{1-\beta}} \rlap{\ .}
\end{equation*}

At first order, the integral of $A$ along the geodesic is needed. Using the same notations as in section~\ref{subsec:order1} where variables have been rescaled such that $\nu = 1$, it gets an additional term
\begin{equation*}
  \cA = \cA_{\mathrm{SABR}} + \frac{\rho \kappa}{\sqrt{1-\rho^2}} \left[ 2 \left( \atan(t_2) - \atan(t_1) \right) - \frac{\overline V}{R} \ln\!\left( \frac{t_2}{t_1}\right) \right] + \kappa \left[ \ln\!\left(\frac{V}{\alpha} \right) + \frac{\overline V}{V} - \frac{\overline V}{\alpha}\right] \rlap{\ .}
\end{equation*}

The second order correction involves a one-dimensional integral which can be computed numerically by a quadrature with a few points, although it may be possible to get an analytical expression.

\end{appendices}

\pagebreak
\bibliographystyle{alpha}
\bibliography{vol}

\end{document}